\documentclass[11pt]{aastex61} 

\usepackage{graphicx}           
\usepackage{latexsym}           
\usepackage{amsmath}            
\usepackage{nccmath}            
\usepackage{bm}                 
\usepackage[english]{babel}
\usepackage{color}              
\usepackage{booktabs}           
\usepackage{multirow}           
\usepackage{natbib}

\newcommand{\del}[1]{\relax}%
\newcommand{\DELETED}[1]{\relax}%
\newcommand{\DEL}[1]{\relax}%
{\relax}%

\definecolor{violet}  {rgb}{1.0,0.0,1.0}

\definecolor{dviolet} {rgb}{0.75,0.0,1.0}

\definecolor{blue}    {rgb}{0.0,0.7,1.0}

\definecolor{lblue}   {rgb}{0.5,1,1}

\definecolor{dblue}   {rgb}{0.0,0.0,1.0}

\definecolor{blgr}    {rgb}{0.70,0.80,1.00}

\definecolor{navy}    {rgb}{0.00,0.00,0.48}

\definecolor{green}   {rgb}{0.7,1.0,0.0}

\definecolor{dgreen}  {rgb}{0.0,1.0,0.0}

\definecolor{lgreen}  {rgb}{0.0,0.8,0.0}

\definecolor{dg}      {rgb}{0.0,0.6,0.0}

\definecolor{orange}  {rgb}{1.0,0.5,0.0}

\definecolor{dorange} {rgb}{1.0,0.6,0.0}

\definecolor{brown}   {rgb}{0.1,0.1,0.0}

\definecolor{lbrown}  {rgb}{0.7,0.5,0.0}

\definecolor{red}     {rgb}{1,0.0,0.0}

\definecolor{dred}    {rgb}{0.6,0.0,0.0}

\definecolor{grey}    {rgb}{0.1,0.1,0.1}

\definecolor{lgrey}   {rgb}{0.5,0.5,0.5}

\definecolor{black}   {rgb}{0.0,0.0,0.0}

\newcommand\n            {\noindent}

\newcommand\bn           {\bigskip\noindent}
\newcommand\mn           {\medskip\noindent}
\newcommand\sn           {\smallskip\noindent}
\newcommand\cl           {\centerline}
\newcommand\ve           {\vfill\eject}

\newcommand\arcspt       {{$\buildrel{\prime\prime}\over .$}}
\newcommand\degree       {{\ifmmode^\circ\else$^\circ$\fi}} 
\newcommand\arcm         {{\ifmmode {'\ }\else$'     $\fi}} 
\newcommand\arcs         {{\ifmmode{''\ }\else$''    $\fi}} 

\newcommand{\bul}        {$\bullet$\ }
\newcommand\cge          {{$_ >\atop{^\sim}$}}
\newcommand\cle          {{$_ <\atop{^\sim}$}}

\newcommand\eg           {{\it e.g.},}
\newcommand\ie           {{\it i.e.},}

\newcommand\kms          {{km\ s$^{-1}$}}

\newcommand\Lya          {{Ly$\alpha$} }

\newcommand\Lo           {{$L_{\odot}$}}

\newcommand\Lbol         {{$L_{\rm bol}$}}

\newcommand\magarc       {{mag\ arcsec$^{-2}$}}
\newcommand\arcsecsq     {{arcsec$^{2}$}}
\newcommand\mum          {{$\mu$m}} 
\newcommand\MBH          {{$M_{\rm BH}$} }

\newcommand\Mo           {{M$_{\odot}$}}

\newcommand\nWsqmsr      {{nW\ m$^{-2}$\ sr$^{-1}$}}
\newcommand\nWsqmsrsq    {{nW$^{2}$\ m$^{-4}$\ sr$^{-2}$}}

\newcommand\Ro           {{R$_{\odot}$}}

\newcommand\rhl          {{$r_{\rm hl}$} }

\newcommand\Rs           {{$R_{\rm s}$}}

\newcommand\RUV          {{$R_{\rm UV}$} }
\newcommand\RMS          {{$R_{\rm MS}$} }

\newcommand\tauMS        {{$\tau_{\rm MS}$}}
\newcommand\tauGB        {{$\tau_{\rm GB}$}}

\newcommand\Teff         {{$T_{\rm eff}$}}

\newcommand\vT           {{$v_{T}$}}

\newcommand\VTs          {{$V_{T},s$}}

\newcommand\Zo           {{$Z_{\odot}$}}

\defcitealias{driver_2016}{D16}
\defcitealias{windhorst_2011}{W11}

\def\ltsima{$\; \buildrel < \over \sim \;$}
\def\lsim{\lower.5ex\hbox{\ltsima}}
\def\gtsima{$\; \buildrel > \over \sim \;$}
\def\gsim{\lower.5ex\hbox{\gtsima}}

\newlength{\txw}
\newlength{\txh}

\begin{document}

\vspace*{-0.500cm}
\n {\Huge Astro2020 Science White Paper} 

\bn 

\bn {\Large On the observability of individual Population III stars and their
stellar-mass black hole accretion disks through cluster caustic transits}

\normalsize

\bn 

\bn \textbf{Thematic Areas:} 
\bul Formation and Evolution of Compact Objects; \bul Cosmology and
Fundamental Physics; \bul Stars and Stellar Evolution; \hspace*{1pt}
\bul Resolved Stellar Populations and their Environments; \bul Galaxy
Evolution; \bul Multi-Messenger Astronomy and Astrophysics. 

\bn 

\bn \textbf{Principal Author:} Rogier A. Windhorst (School of Earth and Space
Exploration, Arizona State University, Tempe, AZ 85287-1404); Email:
Rogier.Windhorst@asu.edu ; Phone: 480-965-7143. 

\bn \textbf{Co-authors:} 
M. Alpaslan (New York U.), 
S. Andrews (U. Western Australia), 
T. Ashcraft (ASU), 
T. Broadhurst (U. Basque Country, Spain), 
D. Coe (STScI), 
S. Cohen (ASU), 
C. Conselice (U. Nottingham, UK), 
J. Diego (Inst. de Fisica de Cantabria, Spain), 
M. Dijkstra (U. Oslo), 
S. Driver (U. Western Australia), 
K. Duncan (U. Leiden, the Netherlands), 
S. Finkelstein (UT Austin), 
B. Frye (U. of Arizona), 
A. Griffiths (U. Nottingham, UK), 
N. Grogin (STScI), 
N. Hathi (STScI), 
A. Hopkins (AAO, Sydney, Australia), 
R. Jansen (ASU), 
B. Joshi (ASU), 
A. Kashlinsky (NASA GSFC), 
W. Keel (U. Alabama), 
P. Kelly (U. Minnesota), 
D. Kim (ASU), 
A. Koekemoer (STScI), 
R. Larson (UT Austin), 
R. Livermore (UT Austin), 
M. Marshall (U. Melbourne, Australia), 
M. Mechtley (ASU), 
N. Pirzkal (STScI), 
M. Rieke (U. of Arizona), 
A. Riess (JHU), 
A. Robotham (U. Western Australia), 
S. Rodney (U. So Carolina), 
H. R\"ottgering (U. Leiden, the Netherlands), 
M. Rutkowski (Minnesota State U), 
R. Ryan Jr. (STScI), 
B. Smith (ASU), 
A. Straughn (NASA GSFC), 
L. Strolger (STScI), 
V. Tilvi (ASU), 
F. Timmes (ASU), 
S. Wilkins (U. Sussex, UK), 
C. Willmer (U. of Arizona), 
R. Windhorst (ASU), 
S. Wyithe (U. Melbourne, Australia), 
H. Yan (U. Missouri), 
A. Zitrin (Ben Gurion U., Israel). 

\del{  

We discuss if Pop III stars and accretion disks around stellar-mass black holes
at z>7 may be directly observable to JWST and the next generation 25-39 meter
ground-based telescopes through cluster caustic transits. About 3-30 of the
best lensing clusters need to be monitored several times a years for several
years to directly detect this. 

}

\bn

\bn {\bf Abstract:}\ 
Recent near-infrared power-spectra and panchromatic Extragalactic Background
Light (EBL) measurements provide upper limits on the integrated near-infrared
surface brightness (SB\cge 31\magarc\ at 2\mum) that may come from Population
III (Pop III) stars and possible accretion disks around resulting stellar-mass
black holes (BHs) in the epoch of First Light, broadly taken from
z$\simeq$7--17. Physical parameters for zero metallicity Pop III stars at z\cge
7 can be estimated from \texttt{MESA} stellar evolution models through
helium-depletion, and for BH accretion disks from quasar microlensing results
and multicolor accretion models. Second-generation non-zero metallicity stars
can form at higher multiplicity, so that BH accretion disks may be fed by
Roche-lobe overflow from lower-mass companions in their AGB stage. The
near-infrared SB constraints can be used to calculate the number of caustic
transits behind lensing clusters that the James Webb Space Telescope (JWST) and
the next generation 25--39 m ground-based telescopes may detect for both Pop
III stars and stellar mass BH accretion disks. Because Pop III stars and
stellar mass BH accretion disks have sizes of a few$\times$10$^{-11}$ arcsec at
z\cge 7, typical caustic magnifications can be $\mu$$\simeq$10$^4$--10$^5$,
with rise times of hours and decline times of \cle 1 year for cluster
transverse velocities of \vT\cle 1000 \kms. Microlensing by intracluster medium
objects can modify transit magnifications, and lengthen visibility times.
Depending on BH masses, accretion-disk radii and feeding efficiencies,
stellar-mass BH accretion-disk caustic transits could outnumber those from Pop
III stars. To observe Pop III caustic transits directly may require monitoring
3--30 lensing clusters to AB\cle 29 mag over a decade or more. {\it Such a
program must be started with JWST at the start of Cycle 1, and --- depending on
the role of microlensing in the Intra Cluster Light (ICL) --- should be
continued for decades with the next generation 25--39 m ground-based
telescopes, where both JWST and the ground-based facilities each will play a
unique and strongly complementary role.}

\ve 


\n {\bf 1. Introduction:}\ In this paper, we consider if the James Webb Space
Telescope \citep[JWST;][]{gardner_2006, rieke_2005, beichman_2012,
windhorst_2008} can detect First Light objects directly. JWST's Near-InfraRed
Camera (NIRCam) is expected to reach medium-deep to deep (AB$\simeq$28.5--29
mag) flux limits routinely, and in ultradeep surveys perhaps as faint as
AB$\simeq$30--31 mag. Unlensed Pop III stars or resulting stellar-mass black
hole (BH) accretion disks at z$\simeq$7--25 likely have fluxes of 
AB$\simeq$35--43 mag, and therefore are not directly detectable by JWST, not
even via ordinary gravitational lensing \citep[\eg][]{rydberg_2013}, which
gives typical magnifications of $\mu$$\simeq$10\ or $\sim$2.5 mag
\citep[\eg][]{lotz_2017}. However, cluster caustic transits, when a compact
restframe UV-source transits a caustic due to the cluster motion in the sky ---
or due to significant velocity substructure within the cluster --- could
magnify such compact objects temporarily by factors of
$\mu$$\simeq$10$^3$--10$^5$ \citep[\eg][]{miralda_escude_1991, zackrisson_2015,
kelly_2017, kelly_2018, diego_2018, rodney_2018, windhorst_2018, chen_2019,
kaurov_2019}. This could temporarily boost the brightness of a very compact
object by $\mu$$\simeq$7.5--12.5 mag. Thus if Pop III stars with
AB$\simeq$35--41.5 mag at redshifts z$\simeq$7--17 --- or accretion disks
around their resulting stellar mass BHs --- are sufficiently numerous in the
sky, they could be detectable for a few months in a medium-deep or deep
(AB$\simeq$28.5--29 mag) monitoring program with JWST of suitable foreground
clusters. 

\mn {\bf 2a. Constraints to the Sky-Surface Brightness from Pop III Stars at
z\cge 7:}\ Before we can estimate the number of cluster caustic transits of Pop
III objects, we must estimate the maximum possible contribution of Pop III stars
and stellar-mass BH accretion disks to the observed near--IR sky surface
brightness (the diffuse EBL from z\cge 7). Based on metallicity arguments,
\citet{madau_2005} suggested that Pop III stars contribute less than a few
\nWsqmsr\ to the (1--4 \mum) InfraRed Background (IRB). \citet{cooray_2012}
estimated the Pop III flux to be \cle 0.04 \nWsqmsr\ based on a detailed Pop III
model for reionization. To confirm these numbers, we estimate the average sky-SB
from star-forming objects at z$\simeq$7--8 from the actual HUDF data corrected
for incompleteness \citep{bouwens_2015}. To this we need to add the light from
the steep faint-end of the galaxy luminosity function (LF) at z\cge 7 from {\it
inferred but unseen} Pop III objects beyond the detection limit of the deepest
HST images, {\it and} add an estimate of the {\it maximum} sky-SB from
z$\simeq$9 to z$\simeq$17 that is not yet observed. For this, we use the
extrapolation of the \citet{madau_2014} cosmic SFR, which is $\sim$0.3 dex
above the fits of \citet{finkelstein_2016} and \citet{madau_2017} to the most
recent WFC3 data at z$\simeq$8--10, resulting in a {\it most conservative upper
limit to the total 2.0 \mum\ sky-SB} from star-forming objects at 7\cle z\cle
17 down to the luminosity of a single Pop III star: SB\cge 31 \magarc. 

\sn {\bf 2b. Diffuse EBL Limits Adopted for Pop III Stellar Mass BH Accretion
Disks:}\ \cite{kashlinsky_2012, kashlinsky_2015}, \citet{cappelluti_2013},
\citet{helgason_2016}, and \citet{mitchell-wynne_2016} provided estimates of the
object-free IR-power spectrum. After carefully subtracting all objects in
ultradeep Spitzer 3.6 and 4.5 \mum\ images in the GOODS-South field
\citep{grogin_2011, koekemoer_2011}, these papers found a consistent rather
uniform {\it signal} in the power-spectrum on 100--1000\arcs\ scales with an
$r.m.s.$ (amplitude)$^2$ of \cle 0.004 \nWsqmsrsq, which is relatively flat on
the angular scales where it is well sampled, and is fairly similar between 3.6
and 4.5 \mum. This 3.5 \mum\ power spectrum amplitude provides an upper limit to
the diffuse 3.5 \mum\ sky-SB that may be generated by objects at z\cge 7.
\citet{cappelluti_2013} cross-correlated the object-subtracted ultradeep Spitzer
images with the deepest object-free 0.2--2 keV Chandra images in the same
CANDELS field, and found a similar signal on \cge 10\arcs\ scales.
\citet{cappelluti_2017} fitted the 0.3--7 keV energy spectrum of the X-ray
background (XRB) with the redshifted X-ray spectra of known populations, and
constrain the fraction of the XRB that can come from unresolved sources ---
possibly early black holes at z\cge 6 --- as\ \cle 3\% of the peak in the
supermassive black hole (SMBH) growth-rate curve at z$\simeq$1--2. If this
Spitzer--Chandra cross-correlation signal is real, the implication is that some
fraction of it may come from First Light objects at z\cge 7. This signal has
also been modeled with Primordial Black Holes \citep[PBHs;][]{kohri_2014},
Direct Collapse Black Holes \citep[DCBHs;][]{yue_2013}, or Obese Black Holes
\citep[OBHs;][]{natarajan_2017} at z\cge 7--8. \citet[][hereafter
W18]{windhorst_2018} adopt the equivalent sky-SB value of \cge 30.8\ \magarc\ at
2.0 \mum\ as the upper limit for BH caustic transit calculations. Note that for
caustic transit calculations it does not matter whether the light that comes
from z\cge 7 exists in {\it faint discrete objects} that have already been
detected down to the HUDF limit, or whether this light is fully {\it
unresolved} below the current HUDF object detection limit of AB$\simeq$30 mag.
Either way, the maximum 2.0 \mum\ SB of $\sim$31\ \magarc\ that can be produced
at z\cge 7 may be subject to cluster caustic transits.

\ve 


\vspace*{-1.00cm}
\n\cl{
\includegraphics[width=0.560\txw,angle=0]{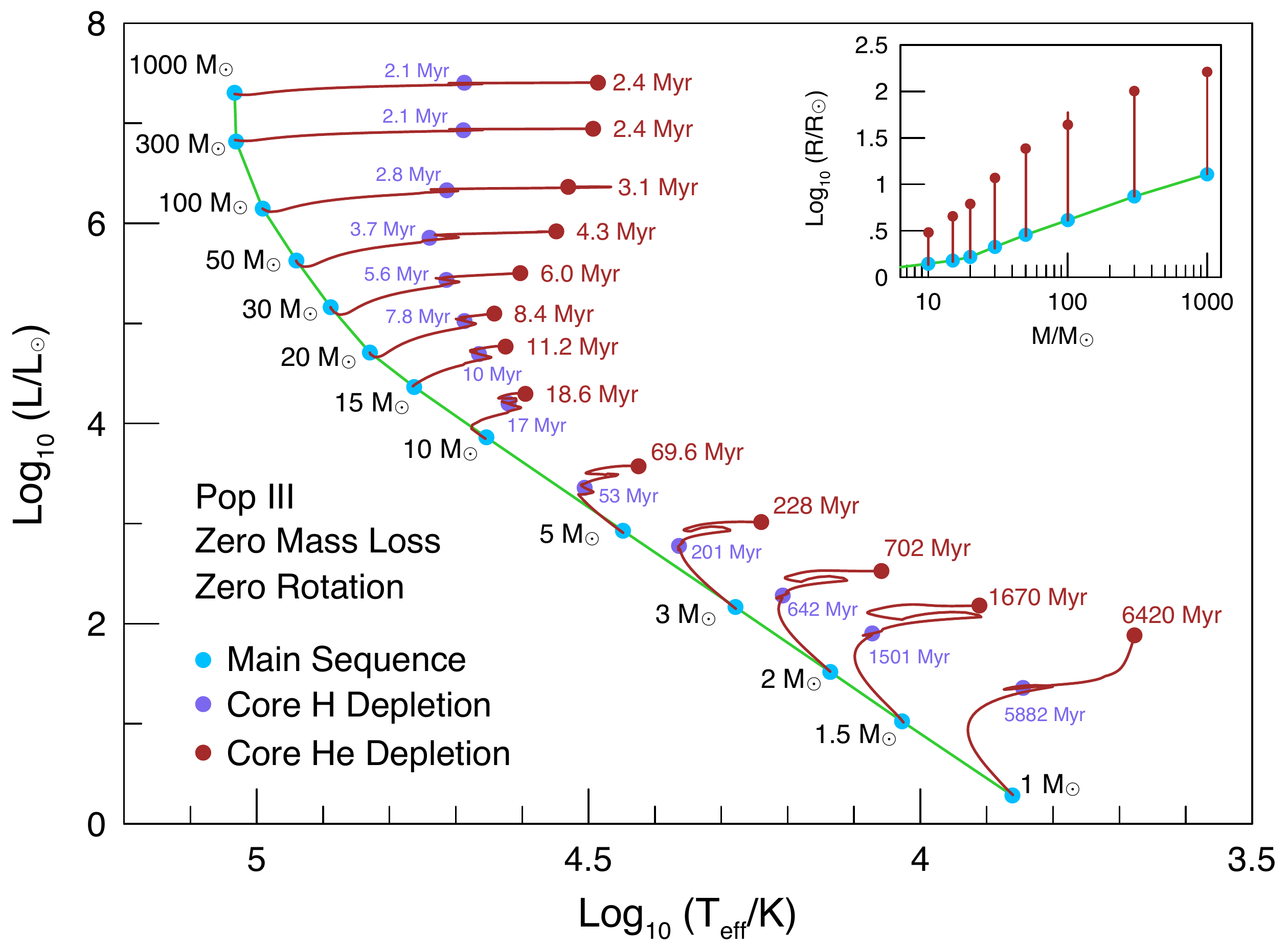}
\includegraphics[width=0.425\txw,angle=0]{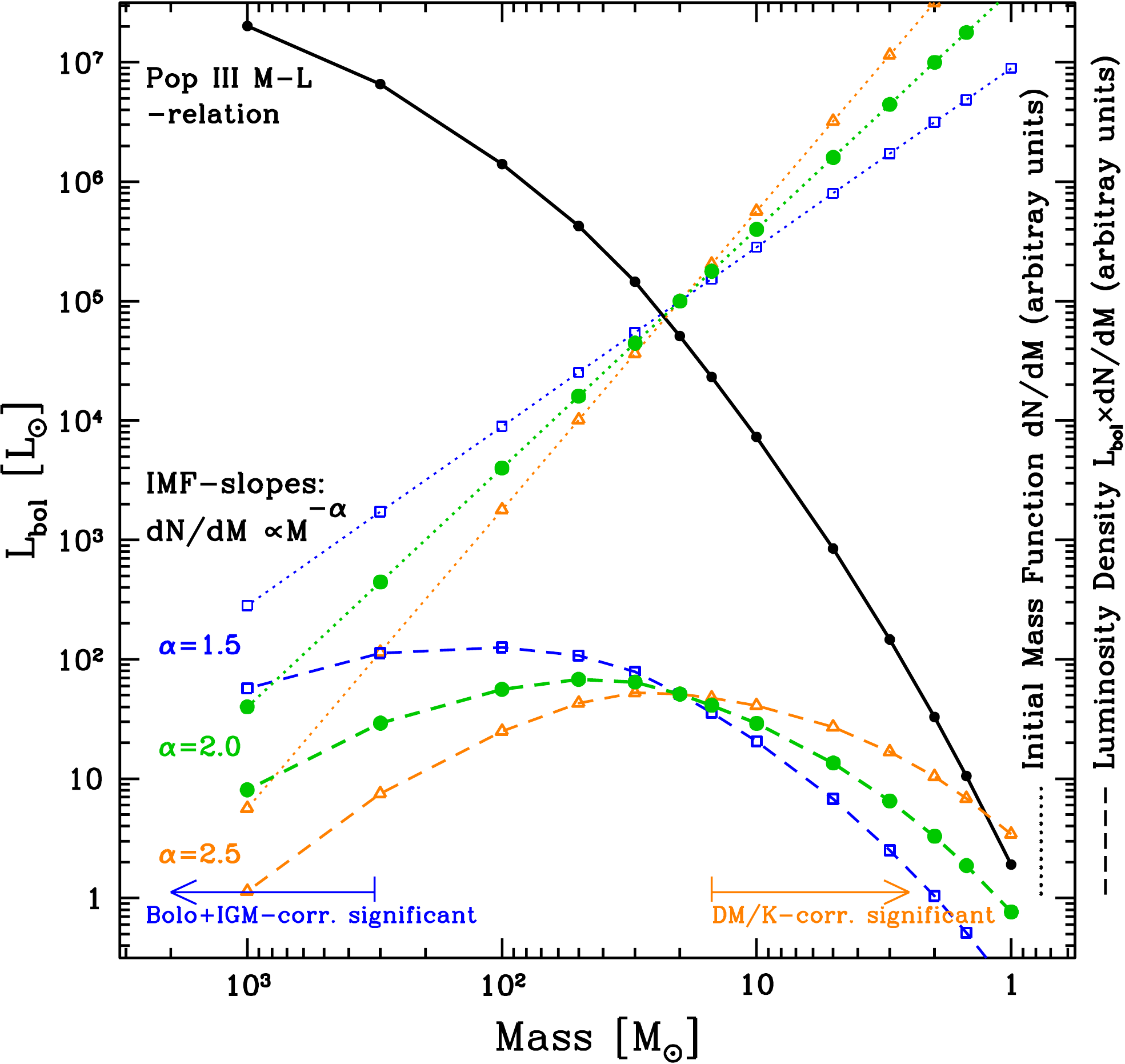}
}

\vspace*{-0.00cm}
\n {\small 
{\bf Fig. 1. (LEFT)}\ HR diagram for non-rotating, zero mass loss, Z = 0.00 \Zo\
\texttt{MESA} models, with evolutionary tracks to core He-depletion shown. Filled
circles, labeled by age, correspond to the models in W18. The inset plot shows
the mass-radius evolution, with filled circles marking the location of ZAMS and
core He-depletion. 
{\bf Fig. 2 (RIGHT):}\ The luminosity density (dashed curves) for early
star-forming objects inferred from the ZAMS Pop III mass-luminosity relation
(solid black line) from W18. The ZAMS Pop III ML-relation is folded with three
different IMF slopes (dotted lines), ranging from $\alpha$=1.5 (top heavy;
blue), $\alpha$=2.0 (normal; green), and $\alpha$=2.5 (steep IMF; orange). For
a Pop III IMF slope of $\alpha$$\simeq$2, the luminosity density peaks around
30 \Mo, while most of the population's luminosity density is produced between
10--100 \Mo, \ie\ the mass range that can produce LIGO-mass BHs!}

\mn {\bf 3a. Physical Parameters Adopted for Pop III Stars from \texttt{MESA}
Models:}\ W18 presented physical properties of Pop III stars from stellar
evolution models with HR-diagrams through the hydrogen-depletion and
helium-depletion stages, and derived their mass-luminosity relation,
bolometric+IGM+K-corrections, and relative contributions to the luminosity
density in a faint star-forming object. These non-rotating, zero metallicity,
zero mass-loss, single 1--1000 \Mo\ star models were calculated using
\texttt{MESA} \citep{paxton_2011, paxton_2013, paxton_2015} with physical and
numerical parameters the same as those in \citet{farmer_2015},
\citet{fields_2016}, and \citet{farmer_2016}. Fig. 1 shows the zero age
main-sequence (ZAMS) in an HR diagram for stellar evolution models with Z =
0.00, and the inset shows their corresponding mass-radius relation. W18 show
that Pop III stars with M$\simeq$30--1000 \Mo\ have ZAMS photospheric
temperatures of 77,000--108,000 K, bolometric luminosities of
\Lbol$\simeq$0.16--20$\times$10$^{6}$ \Lo, stellar radii of \RMS$\simeq$2--13
\Ro, and main sequence (MS) lifetimes of \tauMS$\simeq$2.1--5.6 Myr. They may 
therefore be bright enough for occasional caustic transit detections by JWST, as
calculated by W18. The MS lifetime $\tau$ of the most massive Pop III stars
scales roughly as mass/luminosity. Since luminosities are directly proportional
to ZAMS mass, the \texttt{MESA} models yield MS ages of 5.6--2.1 Myr that are
only weakly dependent on ZAMS mass for the mass range of 30--1000 \Mo. Under the
assumption that (slightly polluted) massive stars at z\cge 7 may occur in binary
or multiple systems, then for a \citet{salpeter_1955} or flatter IMF, stars with
M\cge 30 \Mo\ may have a lower mass companion. Lower mass companion stars with
M\cge 2--5 \Mo\ will be in their RGB--AGB stage for \tauGB\ \cle 30-60 Myr, \ie\
much longer than the plausible lifetime of a massive Pop III primary star. They
could thus be feeding the LIGO-mass BH leftover from the massive Pop III star
after 2.4--6 Myr. As long as the more massive star --- during its short giant
branch (GB) lifetime --- does not transfer the majority of its mass to its
companion star, the resulting BH accretion timescale would be driven by the 
longer GB lifetime of the companion star. 

\sn {\bf 3b. Luminosity Density from Pop III Star Mass-Luminosity Relation and
Initial Mass Function:}\ The ZAMS Pop III mass-luminosity relation in Fig. 2 has
important implications for the mass range that dominates the luminosity density
of a faint star-forming object at z\cge 7. This is indicated in Fig. 2, where
the ZAMS ML-relation is indicated by the solid black line. Three different IMF
slopes are indicated in Fig. 2 (dotted curves), ranging from ``top-heavy''
($\alpha$=1.5; blue), ``intermediate'' ($\alpha$=2.0; green), and ``steep''
($\alpha$=2.5; orange) which bracket a range of plausible IMFs
\citep[\eg][]{bastian_2010, coulter_2017, scalo_1986}. The ZAMS Pop III
ML-relation is folded with these three IMF slopes to yield the luminosity
density in Fig. 2. For an IMF-slope of $\alpha$$\simeq$2, most of the {\it
bolometric energy} from faint star-forming objects at z\cge 7 is thus produced
by Pop III stars with masses between 10--100 \Mo, with a smaller contribution
from stars with M$\simeq$100--1000 \Mo, and a much smaller contribution from
M$\simeq$1--10 \Mo, which is compounded by the significant K-correction for the
lowest mass stars (W18). For an IMF slope of $\alpha$$\simeq$2, the Pop III
luminosity density peaks around 30 \Mo\ with a broad plateau (green dashed
curve in Fig. 2). These are precisely the stars that leave BHs in the mass
range observed by LIGO. 

\mn {\bf 4a. Estimates of Cluster Caustic Transits for Pop III stars:}\ To
estimate the caustic transit rate and duration for Pop III stars, we first need
to evaluate the plausible limits to the transverse velocities of massive lensing
clusters, their typical caustic lengths, and the possible effects from
microlensing. A Pop III caustic-transit observing program with JWST should
select the best lensing clusters with matching prior HST/ACS and WFC3 images,
such as the Hubble Frontier Field clusters \citep[HFF; \eg][]{lotz_2017,
kawamata_2016, lagattuta_2017, acebron_2017, mahler_2018} or the CLASH clusters
\citep[\eg][]{postman_2012, rydberg_2015}. Given the significant differences in
the allowed \vT-values between the three HFF clusters discussed in W18, we adopt
an upper limit of \VTs\cle 1000 \kms. For order-of-magnitude estimates of Pop
III object caustic transits at z\cge 7, we assume average caustic lengths and
geometry. Line integration of the lensing models in clusters like Fig. 3b shows
that the typical {\it total} caustic length is $L_{\rm caust}$\cle 100\arcs,
which we use as upper limit. When a background star crosses a cluster caustic it
can be magnified by a factor of up to $\mu$$\simeq$$10^5$--$10^6$ for a short
period of time (few weeks--months), depending on the strength of the caustic and
the stellar radius, boosting the apparent brightness of the star by
$\sim$12.5--15 mag. Fainter Pop III stars with AB$\simeq$41--43 mag would then
be observable with JWST at AB\cle 28.5--29 mag during such caustic crossing
events. For the more ubiquitous fold caustics, the magnification near a caustic
varies with the distance to the caustic, $d$, as: $\mu = B_o / \sqrt{d}$, \n
where $B_o$ is a constant that depends on the derivatives of the gravitational
potential. For clusters like the HFFs, $B_o$$\simeq$10--20, while $d$ is
expressed in arcseconds \citep[\eg][]{miralda_escude_1991, diego_2018}. Hence,
for a Pop III star at z \cge 7, magnifications of order $\mu$$\simeq$10$^3$ can
be attained once the star is $\simeq$1 pc away from the caustic (or
$d$$\simeq$0\arcspt 001). For an HFF-like cluster with ${\rm L_{\rm
caust}}$$\simeq$100\arcs, this implies that an area of $\sim$0.1 \arcsecsq\ in
the source plane can magnify background stars at z$\simeq$12 by $\mu$\cge 1000,
so that AB\cle 36 mag stars can be lensed to above the detection limit of JWST.
Microlensing by ICL can modify transit magnifications, and lengthen visibility
times.

\sn {\bf 4b. Implied Estimates of Cluster Caustic Transits for Pop III Stars:}\
{\it If} a fraction of the diffuse near-IR background {\it is} generated by Pop
III stars --- with a conservative upper limit to their near-IR sky-SB of \cge
31\ \magarc\ (\S\ 2) --- then what is the probability that JWST will catch a
Pop III star transiting a cluster caustic? We start with the premise that this
maximum 1--4 \mum\ sky-SB results from ZAMS Pop III stars with AB\cge 37.5 mag
at z\cge 7 (\S\ 2). During their RGB and AGB stages, these Pop III stars may
reach AB$\simeq$35 mag at z\cge 7 (W18). Pop III stars in the mass range of
30\cle M\cle 1000 \Mo\ are the most likely to be detected by JWST at z\cge 7 at
AB\cle 28.5--29 mag {\it if} the caustic magnifications reach $\mu$\cge
10$^4$--10$^5$. ZAMS Pop III stars are
$\sim$5.2$\times$10$^{-12}$--7.78$\times$10$^{-11}$ arcsec across at
z$\simeq$12, implying that the brightening time --- defined as the time for the
magnification to go from zero to its maximum value --- is very short
($\sim$0.5--3 hours) when the star transits the caustic {\it starting} at the
``highest-magnification edge''. The star would then stay bright for several
months to a year, with brightness decaying as 1/$\sqrt{t-t_o}$, where ($t-t_o$)
is the time since the stellar disk started the caustic crossing at a time $t_o$.
(A reverse transit may of course also be observed). 

For an IR background of \cge 31 \magarc\ made up of AB$\simeq$41 mag Pop III
stars with M$\simeq$100 \Mo, we estimate that one lensing event can be observed
above a flux limit of AB$\simeq$28.5 mag per cluster per $\sim$2.7 years, or one
event when monitoring $\sim$3 clusters during a year. Because these events
should stay detectable at $\mu>\mu$(M=100\Mo) for $t_{\mu}\simeq 0.4$ years,
this implies that $\sim0.15$ such lensed Pop III sources per cluster would be
observed above the flux limit at any given time. Thus, for 100 \Mo\ Pop III
stars, about 6 clusters observed twice about 6 months apart would make the
likelihood of observing a lensed Pop III star of order unity, while for more
massive stars, detecting a new lensing event (with a time baseline limited to 1
year) would require observation of a larger number of clusters in proportion to
the mass $M$. For lower-mass stars, fewer clusters would need to be observed,
as long as they can appear magnified above the detection thresholds of JWST.
The {\it total} transit rates for stars with M\cge 30 \Mo\ that are in
principle observable with JWST across the caustics are then predicted to be
$\frac{dN_{\rm lens}}{dt}$\cle 0.30 events per cluster per year. Observing more
often when scheduling allows for clusters at higher Zodiacal latitude would be
preferred. Since the uncertainty factors

\ve 


\vspace*{-0.40cm}

\hspace*{-0.40cm}
\begin{minipage}[t]{1.000\txw}
 \begin{minipage}[t]{0.600\txw}
\n \includegraphics[height=0.340\txw,angle=0]{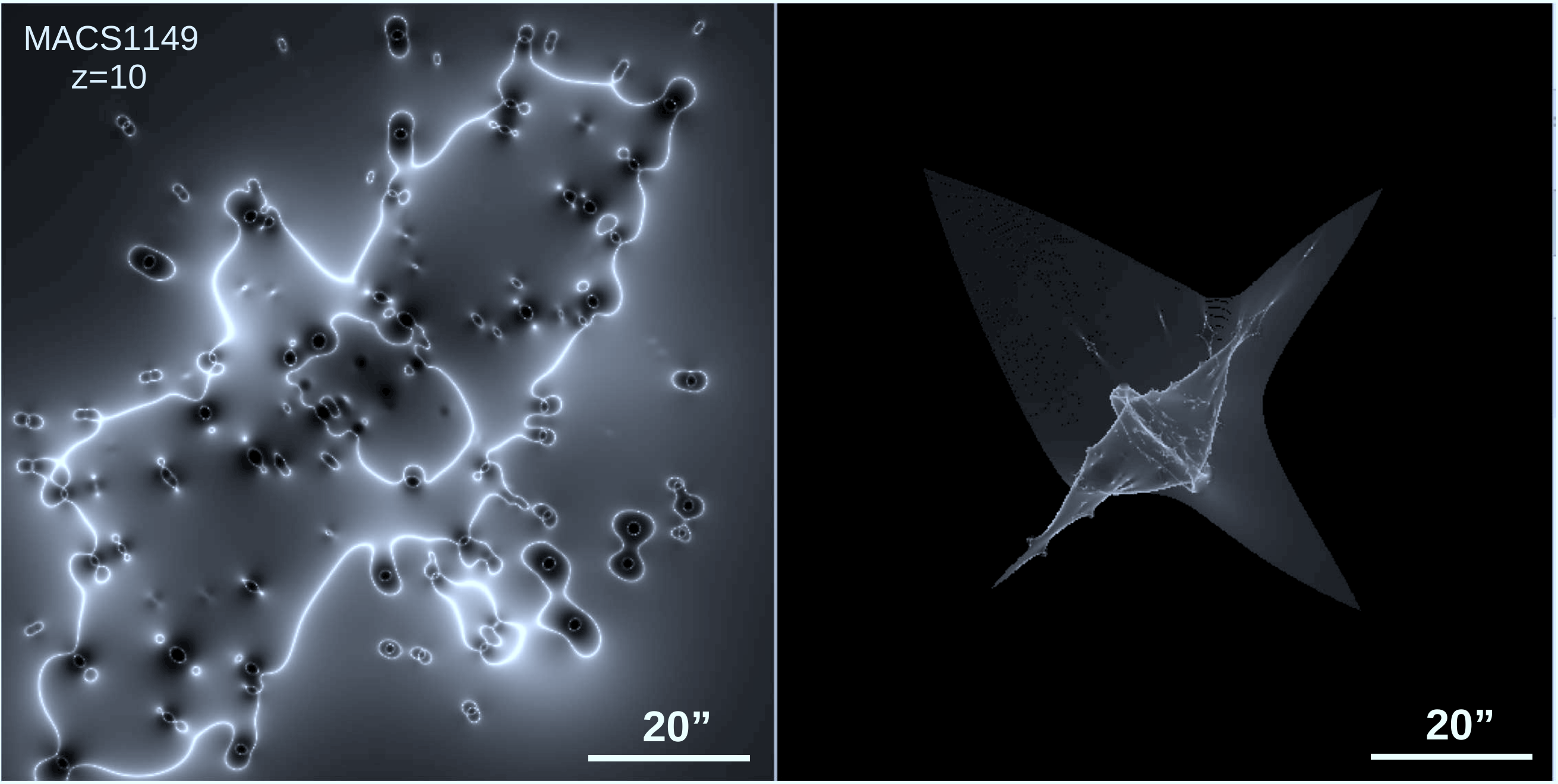} 
 \end{minipage}
\vspace*{-0.00cm}
\hspace*{+1.500cm}
 \begin{minipage}[t]{0.300\txw}
 \vspace*{-5.70cm}
 \hspace*{-0.00cm}
\n {\scriptsize 
{\bf Fig. 3a (LEFT):}\ Lensing magnification map for a galaxy cluster at
z$\simeq$0.54 and a source at z=10 \citep[\eg] []{lotz_2017}. White areas mark
the critical curves, where maximum lensing magnification ($\mu$\cge 10--20) is
observed for a source with \rhl\cle 0\arcspt 5 at z=10. 
{\bf Fig. 3b [RIGHT]:}\ Caustic map produced by the cluster mass model for a
source at z=10. White indicates where a point source at z=10 produces maximum
magnification. The total length of the cluster caustics L\cle 100\arcs\ is the
upper limit used for caustic transits calculations.} 
 \end{minipage}
\end{minipage}

\n in these estimates are $\sim$0.7 dex, JWST may need to monitor 3--30
clusters during its lifetime. Such a survey would need to be maintained until a
sufficient number of Pop III star caustic transits has been detected, so that
the actual Pop III star caustic transit rate can be estimated, and the survey
strategy updated. 

\mn {\bf 5a. Parameters Adopted for Pop III Star Black Hole Accretion Disks:}\
To address under what conditions JWST could detect the UV accretion disks of
Pop III stellar-mass BHs lensed individually through cluster caustic transits
at very high magnification, we first need to discuss their plausible range in
physical properties, and under what conditions these may be fed from early
massive stellar binaries for the expected range in IMF-slope (Fig. 2) and
metallicity evolution \citep{sarmento_2019}. For the Pop III ZAMS mass range in
our \texttt{MESA} models, we adopt similar end-products as in
\citet{woosley_2002}. The recent LIGO detections of stellar-mass BHs at z\cle
0.1 \citep{abbott_2016_a, abbott_2016_c} are very plausibly examples of merging
black hole pairs with M$\simeq$29--36 \Mo, 14--21 \Mo, and 19--31 \Mo,
respectively, about 1--3 Gyr ago \citep{abbott_2016_b, abbott_2016_d,
abbott_2016_e, abbott_2017_a}. For the calculations of Pop III BH accretion
disk caustic transits, we assume that Pop III stars with M\cge 30 \Mo\ --- with
the exception of the mass range of $\sim$100\cle M\cle 200 \Mo\ --- can and
will produce BHs of roughly 15--70\% of the ZAMS Pop III stellar mass, or
M$\simeq$5--720 \Mo\ \citep {woosley_2002}. The Schwarzschild radii of these
Pop III BHs will be in the range \Rs$\simeq$15--2200 km. What matters for the
current work is that, while some massive stars with zero or very low
metallicity may still exist at z$\simeq$7, at the same time a sufficient
fraction of polluted stars (Z\cge 10$^{-4}$ \Zo) already exists at
z$\simeq$7--17 \citep{trenti_2007, sarmento_2018}. The latter are critical,
since they likely formed with a significant fraction of binaries, and so play an
essential role in BH accretion disk feeding via Roche-lobe overflow during
post-main sequence evolution. Any BHs left over after a massive Pop III star's
death may accrete from a surrounding lower-mass, low-metallicity star filling
its Roche lobe during its post-main sequence evolution, causing a UV-bright
accretion disk. The accretion time scales onto these BHs in stellar binaries
are not well known, but may have plausibly lasted as long as the GB lifetimes
of the less massive star in a binary when it fills its Roche lobe (W18).
\citet{blackburne_2011} suggest from their QSO microlensing data that for
\MBH\cge\ 10$^9$\Mo, the accretion disk half-light radii scales as: $r_{\rm hl}
\propto M_{\rm BH}\ ^{\rho}$. Using $\rho$$\simeq$0.5, we obtain consistent BH
UV half-light radii in the range \RUV$\simeq$1--16 \Ro\ for \MBH$\simeq$5--720
\Mo. The bolometric luminosities are 4$\times$10$^4$--6$\times$10$^6$ \Lo\ for
\MBH$\simeq$5--720 \Mo. We confirm these results with multi-color thin
accretion-disk models, where the temperature increases with radius as $T \propto
r^{-3/4}$. Gas on the innermost stable orbit at R$\simeq$3\Rs\ has a maximum
temperature of about $T_{max}$$\simeq$10\ $(\frac{M_{BH}}{100\
M_\odot})^{-\tau}\ keV$ with $\tau$$\simeq$3/8. At these r$_{hl}$-values, the
accretion disks have an effective temperature of \Teff$\simeq$47,500--48,000 K
for M$\simeq$5--720 \Mo. The inner stellar-mass BH accretion disks {\it will}
be significantly hotter than the typical T$\simeq$10$^5$ K temperatures of Pop
III stars, plausibly reaching X-ray temperatures at the innermost radii, and
reaching $\sim$30,000 K at the outermost radii. UV-bright accretion disks ---
if unobscured by surrounding dust --- have SEDs at $\lambda$\cge 1216\AA\ that
can make it past the neutral IGM at z\cge 7 with UV radii \cle 40,000 \Rs.
Their restframe UV-radii are \RUV$\simeq$1--30\ \Ro, and their UV-luminosities
are at most 3$\times$10$^4$--7$\times$10$^6$ \Lo\ for M$_{BH}$$\simeq$5--720
\Mo, respectively. Pop III stellar-mass BH accretion disk radii may thus be
similar to, or somewhat larger than the 1--13 \Ro\ radii of the ZAMS Pop III
stars, but no larger than the Pop III RGB- or AGB-star radii in W18. They would
fit well within the $\sim$7--55 \Ro\ typical Roche lobe sizes seen in massive
binaries observed in our own Galaxy, and so are can be fed from a less massive
RGB/AGB star in the binary that is filling its Roche lobe. 

\sn {\bf 5b. Estimates of Caustic Transits for Pop III Star Black Hole Accretion
Disks:}\ Pop III stars with 30\cle M\cle 1000 \Mo\ that produce BHs have ZAMS
ages of 5.6--2.1 Myr (Fig. 1) with an average of $\sim$3 Myr. Pop III stars of
masses M$\simeq$2--20 \Mo\ live considerably longer than this during their AGB
stage, where they could fill their Roche lobes for up to 0.6--60 Myr, with an
average GB age of $\sim$6 Myr. Hence, during their AGB stage 2--20 \Mo\ stars
could feed the BH that is left by a 30--1000 \Mo\ star for a maximum duration
that is significantly longer than the ZAMS lifetime of that massive Pop III star
(Fig. 1). Depending on how steady and efficient BH feeding by a lower mass AGB
star in its Roche lobe is, stellar-mass BH accretion disks may be about as
likely as Pop III stars at z\cge 7 to cause cluster caustic transits that could
be observed by JWST, and possibly more likely. Stellar-mass BH accretion disks
with a SB$\simeq$31 \magarc\ (or $\sim$1 \nWsqmsr) could produce about one
caustic transit per 5 clusters per year, and perhaps as many as one event per 2
clusters per year. A dedicated JWST program that monitors 3 clusters per year
for a number of years could thus detect several caustic transits for Pop III
stellar-mass BH accretion disks. 

\mn {\bf 6. Possible Observing Programs to Detect Pop III Caustic Transits:}\ To
observe caustic transits from First Light objects, a dedicated JWST observing
program will be required of at least several, and perhaps up to 30 clusters for
a duration of 1--10 years. Depending on their exact contribution to the {\it
diffuse} 1--4 \mum\ sky-SB (\cle 0.01--0.1 \nWsqmsr), such a JWST observing
program to detect individual Pop III stars and/or stellar-mass BH accretion
disks at z\cge 7 may well require to monitor --- in the optimistic case that
{\it most} of the NIR power-spectrum signal comes from z\cge 7 --- a few
suitable galaxy clusters a number of times during a year. All of these cluster
observations would require coeval images in four NIRCam filter-pairs and/or
four NIRISS filters to constrain the spectral signature and redshift of a Pop
III caustic transit candidate. The caustic transits would appear as z\cge 7
dropout candidates that vary with time, either increasing rapidly and then
slowly fading, or vice versa. The one significant difference between Pop III
stellar-mass BH accretion disks and Pop III stars is likely the presence of a
hard X-ray component that contributes very significantly at the inner accretion
disk radii, and that will {\it also} have a significant energy tail longwards
of \Lya 1216 \AA. No such X-ray component would exist for Pop III stars, since
their stellar photospheres have nearly uniform temperatures of T$\simeq$10$^5$
K (Fig. 1). Hence, Pop III stars will not show chromatic behavior that may be
traced during a caustic transit, but BH accretion disks could show such
chromaticity if they were detected close to the actual caustic transit. 

The next generation 25--40 m ground-based telescopes --- the European Extremely
Large Telescope (E-ELT), the Giant Magellan Telescope (GMT), and the Thirty
Meter Telescope (TMT) --- will have much larger collecting area, and narrower
PSFs when using Multi-Conjugate (laser-assisted) Adaptive Optics, although
perhaps not as stable as JWST's PSFs, and they will have much lower Strehl
ratios. They will also have a 1--2 \mum\ sky foreground that is \cge 7 mag
brighter than JWST's in L2. As a consequence, the next generation ground-based
telescopes may be able to reach AB\cle 29 mag in integrations of hours at 1--2
\mum, but --- given their adaptive optics --- only over a smaller FOV (\cle
20\arcs$\times$20\arcs---1\arcm$\times$1\arcm). Ground-based telescopes will
have reduced sensitivity at wavelengths $\lambda$\cge 2--2.2 \mum\ because of
the strongly increasing thermal foreground. For that reason, JWST will be able
to better address any chromatic differences between caustic transits of Pop III
stars and their stellar-mass BH accretion-disks, especially those at z \cge 12
that require several very sensitive filters at $\lambda$\cge 2 \mum, where
ground-based telescopes cannot reach AB$\sim$29 mag due to the much brighter
thermal foreground. Confirming spectra of caustic transits by Pop III stars or
their stellar-mass BH accretion disks could be taken with the JWST NIRISS and
NIRSpec spectrographs. In summary, the next generation ground-based telescopes
can monitor at 1--2 \mum\ --- over a much longer period than JWST --- individual
Pop III caustic transits that JWST will have detected at 1--5 \mum\ during its
lifetime, and also discover new ones on timescales longer than JWST's lifetime.
This capability would be particularly useful to follow-up on caustic transits
that may be affected by microlensing, and so may stretch out over many decades.
In conclusion, unlensed Pop III stars or stellar-mass BH accretion disks may
have fluxes of AB$\simeq$35--41.5 mag at z$\simeq$7--17, and so will {\it not}
be directly detectable by JWST. However, cluster caustic transits with
magnifications of $\mu$$\simeq$10$^4$--10$^5$ may well render them temporarily
detectable to JWST in medium-deep to deep observations (AB\cle 28.5--29 mag) on
timescales of months to a year, with rise-times less than a few hours. 



\ve 


\bibliographystyle{aasjournal}

\begin{thebibliography}{}

\expandafter\ifx\csname natexlab\endcsname\relax\def\natexlab#1{#1}\fi

\providecommand{\url}[1]{\href{#1}{#1}}

\bibitem[{{Abbott} {et~al.}(2016{\natexlab{a}}){Abbott}, {Abbott}, {Abbott},
{Abernathy}, {Acernese}, {Ackley}, {Adams}, {Adams}, {Addesso}, {Adhikari}, \&
et~al.}]{abbott_2016_a} {Abbott}, B.~P., {Abbott}, R., {Abbott}, T.~D., {et~al.}
2016{\natexlab{a}}, Physical Review Letters, 116, 061102

\bibitem[{{Abbott} {et~al.}(2016{\natexlab{b}}){Abbott}, {Abbott}, {Abbott},
{Abernathy}, {Acernese}, {Ackley}, {Adams}, {Adams}, {Addesso}, {Adhikari}, \&
et~al.}]{abbott_2016_b} ---. 2016{\natexlab{b}}, Physical Review Letters, 116,
241103

\bibitem[{{Abbott} {et~al.}(2016{\natexlab{c}}){Abbott}, {Abbott}, {Abbott},
{Abernathy}, {Acernese}, {Ackley}, {Adams}, {Adams}, {Addesso}, {Adhikari}, \&
et~al.}]{abbott_2016_c} ---. 2016{\natexlab{c}}, \apjl, 818, L22

\bibitem[{{Abbott} {et~al.}(2016{\natexlab{d}}){Abbott}, {Abbott}, {Abbott},
{Abernathy}, {Acernese}, {Ackley}, {Adams}, {Adams}, {Addesso}, {Adhikari}, \&
et~al.}]{abbott_2016_d} ---. 2016{\natexlab{d}}, \apjl, 832, L21

\bibitem[{{Abbott} {et~al.}(2016{\natexlab{e}}){Abbott}, {Abbott}, {Abbott},
{Abernathy}, {Acernese}, {Ackley}, {Adams}, {Adams}, {Addesso}, {Adhikari}, \&
et~al.}]{abbott_2016_e} ---. 2016{\natexlab{e}}, \apjl, 833, L1

\bibitem[Abbott et al.(2017{\natexlab{a}})]{abbott_2017_a} Abbott, B.~P.,
Abbott, R., Abbott, T.~D., et al.\ 2017{\natexlab{a}}, Physical Review Letters,
118, 221101

\bibitem[Abbott et al.(2017{\natexlab{b}})]{abbott_2017_b} Abbott, B.~P.,
Abbott, R., Abbott, T.~D., et al.\ 2017{\natexlab{b}}, Physical Review Letters,
119, 161101 

\bibitem[Abbott et al.(2017{\natexlab{c}})]{abbott_2017_c} Abbott, B.~P.,
Abbott, R., Abbott, T.~D., et al.\ 2017{\natexlab{c}}, \apjl, 848, L13 

\bibitem[Abel et al.(2002)]{abel_2002} Abel, T., Bryan, G.~L., \& Norman,
M.~L.\ 2002, Science, 295, 93

\bibitem[Acebron et al.(2017)]{acebron_2017} Acebron, A., Jullo, E., Limousin,
M., et al.\ 2017, \mnras, 470, 1809 

\bibitem[Adams et al.(2006)]{adams_2006} Adams, F.~C., Proszkow, E.~M.,
Fatuzzo, M., \& Myers, P.~C.\ 2006, \apj, 641, 504 

\bibitem[Adams(2010)]{adams_2010} Adams, F.~C.\ 2010, \araa, 48, 47 

\bibitem[Ahnen et al.(2016)]{ahnen_2016} Ahnen, M.~L., Ansoldi, S., Antonelli,
L.~A., et al.\ 2016, \aap, 590, A24 

\bibitem[Alpaslan et al.(2012)]{alpaslan_2012} Alpaslan, M., Robotham, A.~S.~G.,
Driver, S., et al.\ 2012, \mnras, 426, 2832 

\bibitem[Andrews et al.(2017{\natexlab{a}})]{andrews_2017_a} Andrews, S.~K.,
Driver, S.~P., Davies, L.~J.~M., et al.\ 2017{\natexlab{a}}, \mnras, 464, 1569 

\bibitem[Andrews et al.(2017{\natexlab{b}})]{andrews_2017_b} Andrews, S.~K.,
Driver, S.~P., Davies, L.~J.~M., et al.\ 2017{\natexlab{b}}, \mnras, 470, 1342 

\bibitem[Angus \& McGaugh(2008)]{angus_2008} Angus, G.~W., \& McGaugh, S.~S.\
2008, \mnras, 383, 417 

\bibitem[{{Arendt} {et~al.}(2016){Arendt}, {Kashlinsky}, {Moseley}, \&
{Mather}}]{arendt_2016} {Arendt}, R.~G., {Kashlinsky}, A., {Moseley}, S.~H., \&
{Mather}, J. 2016, \apj, 824, 26

\bibitem[Ashcraft et al.(2018)]{ashcraft_2018} Ashcraft, T.~A., Windhorst,
R.~A., Jansen, R.~A., et al.\ 2018, \pasp, 130, 064102 

\bibitem[Badenes et al.(2018)]{badenes_2018} Badenes, C., Mazzola, C.,
Thompson, T.~A., et al.\ 2018, \apj, 854, 147

\bibitem[Bahcall \& Oh(1996)]{bahcall_1996} Bahcall, N.~A., \& Oh, S.~P.\
1996, \apjl, 462, L49 

\bibitem[Barkana \& Loeb(2001)]{barkana_2001} Barkana, R., \& Loeb, A.\
2001, \physrep, 349, 125

\bibitem[{{Barkana} \& {Loeb}(2002)}]{barkana_2002} {Barkana}, R., \& {Loeb}, A.
2002, \apj, 578, 1

\bibitem[{{Barkat} {et~al.}(1967){Barkat}, {Rakavy}, \& {Sack}}]{barkat_1967}
{Barkat}, Z., {Rakavy}, G., \& {Sack}, N. 1967, Physical Review Letters, 18, 379

\bibitem[Bastian et al.(2010)]{bastian_2010} Bastian, N., Covey, K.~R., \&
Meyer, M.~R.\ 2010, \araa, 48, 339 

\bibitem[{{Beichman} {et~al.}(2012){Beichman}, {Rieke}, {Eisenstein}, {Greene},
{Krist}, {McCarthy}, {Meyer}, \& {Stansberry}}]{beichman_2012} {Beichman},
C.~A., {Rieke}, M., {Eisenstein}, D., {et~al.} 2012, \procspie, 8442, Space
Telescopes and Instrumentation: Optical, Infrared, \& Millimeter Wave, 84422N 

\bibitem[{{Belczynski} {et~al.}(2016){Belczynski}, {Heger}, {Gladysz}, {Ruiter},
{Woosley}, {Wiktorowicz}, {Chen}, {Bulik}, {O'Shaughnessy}, {Holz}, {Fryer}, \&
{Berti}}]{belczynski_2016} {Belczynski}, K., {Heger}, A., {Gladysz}, W.,
{et~al.} 2016, \aap, 594, A97

\bibitem[Bertin \& Arnouts(1996)]{bertin_2006} Bertin, E., \& Arnouts, S.\ 1996,
\aaps, 117, 393 

\bibitem[{{Bessell} {et~al.}(1998){Bessell}, {Castelli}, \&
{Plez}}]{bessell_1998} {Bessell}, M.~S., {Castelli}, F., \& {Plez}, B. 1998,
\aap, 333, 231

\bibitem[Biteau \& Williams(2015)]{biteau_2015} Biteau, J., \& Williams, D.~A.\
2015, \apj, 812, 60 

\bibitem[Blackburne et al.(2011)]{blackburne_2011} Blackburne, J.~A., Pooley,
D., Rappaport, S., \& Schechter, P.~L.\ 2011, \apj, 729, 34 

\bibitem[{{Bond} {et~al.}(1984){Bond}, {Arnett}, \& {Carr}}]{bond_1984} {Bond},
J.~R., {Arnett}, W.~D., \& {Carr}, B.~J. 1984, \apj, 280, 825

\bibitem[Bouwens et al.(2015)]{bouwens_2015} Bouwens, R.~J., Illingworth,
G.~D., Oesch, P.~A., et al.\ 2015, \apj, 803, 34 

\bibitem[Bouwens et al.(2017)]{bouwens_2017} Bouwens, R.~J., Illingworth,
G.~D., Oesch, P.~A., et al.\ 2017, \apj, 843, 41 

\bibitem[Bovill(2016)]{bovill_2016} Bovill, M. S.\ 2016, presentation at the
October 2016 Montreal JWST Workshop
\url{http://craq-astro.ca/jwst2016/agenda_en.php/}


\bibitem[Bromm, Kudritzki, \& Loeb(2001)]{bromm_2001} Bromm, V., Kudritzki,
R.P., \& Loeb, A.\ 2001, \apj, 552, 464

\bibitem[Butler \& Bloom(2011)]{butler_2011} Butler, N.~R., \& Bloom,
J.~S.\ 2011, \aj, 141, 93 

\bibitem[Caminha et al.(2017)]{caminha_2017} Caminha, G.~B., Grillo, C., Rosati,
P., et al.\ 2017, \aap, 600, A90 

\bibitem[Calzetti et al.(1994)]{calzetti_1994} Calzetti, D., Kinney, A.~L., \&
Storchi-Bergmann, T.\ 1994, \apj, 429, 582 

\bibitem[Cannizzo et al.(1988)]{cannizzo_1988} Cannizzo, J.~K., Shafter,
A.~W., \& Wheeler, J.~C.\ 1988, \apj, 333, 227 

\bibitem[{{Cappelluti} {et~al.}(2013){Cappelluti}, {Kashlinsky}, {Arendt},
{Comastri}, {Fazio}, {Finoguenov}, {Hasinger}, {Mather}, {Miyaji}, \&
{Moseley}}]{cappelluti_2013} {Cappelluti}, N., {Kashlinsky}, A., {Arendt},
R.~G., {et~al.} 2013, \apj, 769, 68

\bibitem[Cappelluti et al.(2017)]{cappelluti_2017} Cappelluti, N., Li, Y.,
Ricarte, A., et al.\ 2017, \apj, 837, 19 

\bibitem[{{Casagrande} {et~al.}(2006){Casagrande}, {Portinari}, \&
{Flynn}}]{casagrande_2006} {Casagrande}, L., {Portinari}, L., \& {Flynn}, C.
2006, \mnras, 373, 13

\bibitem[Castor et al.(1975)]{castor_1975} Castor, J.~I., Abbott, D.~C., \&
Klein, R.~I.\ 1975, \apj, 195, 157 

\bibitem[{{Chatzopoulos} {et~al.}(2013){Chatzopoulos}, {Wheeler}, \&
{Couch}}]{chatzopoulos_2013} {Chatzopoulos}, E., {Wheeler}, J.~C., \& {Couch},
S.~M. 2013, \apj, 776, 129

\bibitem[Chen et al.(2019)]{chen_2019} Chen, W., Kelly, P.~L., Diego, J.~M., 
et al.\ 2019, astro-ph/1902.05510

\bibitem[Choi et al.(2016)]{choi_2016} Choi, J., Dotter, A., Conroy,
C., et al.\ 2016, \apj, 823, 102 

\bibitem[Chornock et al.(2017)]{chornock_2017} Chornock, R., Berger, E.,
Kasen, D., et al.\ 2017, \apjl, 848, L19 

\bibitem[Clowe et al.(2006)]{clowe_2006} Clowe, D., Brada{\v c}, M., Gonzalez,
A.~H., et al.\ 2006, \apjl, 648, L109 

\bibitem[Cohen et al.(2006)]{cohen_2006} Cohen, S.~H., Ryan, R.~E., Jr.,
Straughn, A.~N., et al.\ 2006, \apj, 639, 731 

\bibitem[Conroy(2013)]{conroy_2013} Conroy, C.\ 2013, \araa, 51, 393 

\bibitem[{{Cooray} {et~al.}(2012){Cooray}, {Gong}, {Smidt}, \&
{Santos}}]{cooray_2012} {Cooray}, A., {Gong}, Y., {Smidt}, J., \& {Santos},
M.~G. 2012, \apj, 756, 92

\bibitem[Cowperthwaite et al.(2017)]{cowperthwaite_2017} Cowperthwaite, P.~S.,
Berger, E., Villar, V.~A., et al.\ 2017, \apjl, 848, L17 

\bibitem[Coulter et al.(2017)]{coulter_2017} Coulter, D.~A., Lehmer, B.~D., 
Eufrasio, R.~T., et al.\ 2017, \apj, 835, 183 

\bibitem[{{de Mink} \& {Mandel}(2016)}]{de-mink_2016} {de Mink}, S.~E., \&
{Mandel}, I. 2016, \mnras, 460, 3545

\bibitem[Diaferio(1999)]{diaferio_1999} Diaferio, A.\ 1999, \mnras, 309, 610 

\bibitem[Diego et al.(2015{\natexlab{a}})]{diego_2015_a} Diego, J.~M.,
Broadhurst, T., Molnar, S.~M., Lam, D., \& Lim, J.\ 2015{\natexlab{a}}, \mnras,
447, 3130 

\bibitem[Diego et al.(2015{\natexlab{b}})]{diego_2015_b} Diego, J.~M.,
Broadhurst, T., Zitrin, A., et al.\ 2015{\natexlab{b}}, \mnras, 451, 3920 

\bibitem[Diego et al.(2016{\natexlab{a}})]{diego_2016_a} Diego, J.~M.,
Broadhurst, T., Chen, C., et al.\ 2016{\natexlab{a}}, \mnras, 456, 356 

\bibitem[Diego et al.(2016{\natexlab{b}})]{diego_2016_b} Diego, J.~M.,
Broadhurst, T., Wong, J., et al.\ 2016{\natexlab{b}}, \mnras, 459, 3447 

\bibitem[Diego et al.(2018)]{diego_2018} Diego, J.~M., Kaiser, N.,
Broadhurst, T., et al.\ 2018, \apj, 857, 25 

\bibitem[Dressler(1991)]{dressler_1991} Dressler, A.\ 1991, \nat, 350, 391 

\bibitem[{{Driver} {et~al.}(2016){Driver}, {Andrews}, {Davies}, {Robotham},
{Wright}, {Windhorst}, {Cohen}, {Emig}, {Jansen}, \& {Dunne}}]{driver_2016}
{Driver}, S.~P., {Andrews}, S.~K., {Davies}, L.~J., {Robotham}, A.~S.~G.,
{Wright}, A.~H., {Windhorst}, R.~A., {Cohen}, S.~H., {Emig}, K., {Jansen}, R.~A.
\& {Dunne}, L. 2016, \apj, 827, 108 (D16)

\bibitem[{{Duch{\^e}ne} \& {Kraus}(2013)}]{duchene_2013} {Duch{\^e}ne}, G., \&
{Kraus}, A. 2013, \araa, 51, 269

\bibitem[Dwek \& Krennrich(2013)]{dwek_2013} Dwek, E., \& Krennrich, F.\ 2013,
Astroparticle Physics, 43, 112 

\bibitem[Ebeling et al.(2014)]{ebeling_2014} Ebeling, H., Ma, C.-J., \& Barrett,
E.\ 2014, \apjs, 211, 21 

\bibitem[Emilio et al.(2012)]{emilio_2012} Emilio, M., Kuhn, J.~R., Bush, R.~I.,
\& Scholl, I.~F.\ 2012, \apj, 750, 135 

\bibitem[Fan et al.(2001)]{fan_2001} Fan, X., Narayanan, V.~K., Lupton, R.~H.,
et al.\ 2001, \aj, 122, 2833 

\bibitem[Fan et al.(2003)]{fan_2003} Fan, X., Strauss, M.~A., Schneider, D.~P.,
et al.\ 2003, \aj, 125, 1649 

\bibitem[{{Farmer} {et~al.}(2015){Farmer}, {Fields}, \& {Timmes}}]{farmer_2015}
{Farmer}, R., {Fields}, C.~E., \& {Timmes}, F.~X. 2015, \apj, 807, 184

\bibitem[{{Farmer} {et~al.}(2016){Farmer}, {Fields}, {Petermann}, {Dessart},
{Cantiello}, {Paxton}, \& {Timmes}}]{farmer_2016} {Farmer}, R., {Fields}, C.~E.,
{Petermann}, I., {et~al.} 2016, \apjs, 227, 22

\bibitem[Faulkner(1967)]{faulkner_1967} Faulkner, J.\ 1967, \apj, 147, 617

\bibitem[{{Fields} {et~al.}(2016){Fields}, {Farmer}, {Petermann}, {Iliadis}, \&
{Timmes}}]{fields_2016} {Fields}, C.~E., {Farmer}, R., {Petermann}, I.,
{Iliadis}, C., \& {Timmes}, F.~X. 2016, \apj, 823, 46

\bibitem[Fields et al.(2018)]{fields_2018} Fields, C.~E., Timmes, F.~X.,
Farmer, R., et al.\ 2018, \apjs, 234, 19 

\bibitem[Finkelstein et al.(2015)]{finkelstein_2015} Finkelstein, S.~L., Ryan,
R.~E., Jr., Papovich, C., et al.\ 2015, \apj, 810, 71 

\bibitem[Finkelstein(2016)]{finkelstein_2016} Finkelstein, S.~L.\ 2016, \pasa,
33, e037 

\bibitem[Fixsen et al.(1996)]{fixsen_1996} Fixsen, D.~J., Cheng, E.~S., Gales,
J.~M., et al.\ 1996, \apj, 473, 576 

\bibitem[Fixsen(2009)]{fixsen_2009} Fixsen, D.~J.\ 2009, \apj, 707, 916 

\bibitem[Flower(1996)]{flower_1996} Flower, P.~J.\ 1996, \apj, 469, 355 

\bibitem[{{Fraley}(1968)}]{fraley_1968} {Fraley}, G.~S. 1968, \apss, 2, 96

\bibitem[Frank et al.(2002)]{frank_2002} Frank, J., King, A., \& Raine, D.~J.\
2002, Accretion Power in Astrophysics, pp.~398.~ISBN 0521620538 Cambridge
University Press, (Cambridge, UK)

\bibitem[{{Fryer} {et~al.}(2001){Fryer}, {Woosley}, \& {Heger}}]{fryer_2001}
{Fryer}, C.~L., {Woosley}, S.~E., \& {Heger}, A. 2001, \apj, 550, 372

\bibitem[{{Gardner} {et~al.}(2006){Gardner}, {Mather}, {Clampin}, {Doyon},
{Greenhouse}, {Hammel}, {Hutchings}, {Jakobsen}, {Lilly}, {Long}, {Lunine},
{McCaughrean}, {Mountain}, {Nella}, {Rieke}, {Rieke}, {Rix}, {Smith},
{Sonneborn}, {Stiavelli}, {Stockman}, {Windhorst}, \& {Wright}}]{gardner_2006}
{Gardner}, J.~P., {Mather}, J.~C., {Clampin}, M., {et~al.} 2006, \ssr, 123, 485

\bibitem[Giavalisco et al.(2004)]{giavalisco_2004} Giavalisco, M., Ferguson,
H.~C., Koekemoer, A.~M., et al.\ 2004, \apjl, 600, L93 

\bibitem[G{\"o}tberg et al.(2017)]{gotberg_2017} G{\"o}tberg, Y., de
Mink, S.~E., \& Groh, J.~H.\ 2017, \aap, 608, A11 

\bibitem[Greif et al.(2011)]{greif_2011} Greif, T.~H., Springel, V., White,
S.~D.~M., et al.\ 2011, \apj, 737, 75 

\bibitem[Griffiths et al.(2018)]{griffiths_2018} Griffiths, A., Conselice, C.
Conselice, C.~J., Alpaslan, M., et al.\ 2018, MNRAS, 475, 2853 

\bibitem[Grogin et al.(2011)]{grogin_2011} Grogin, N.~A., Kocevski, D.~D.,
Faber, S.~M., et al.\ 2011, \apjs, 197, 35 

\bibitem[Guszejnov et al.(2016)]{guszejnov_2016} Guszejnov, D., Krumholz, M.~R.,
\& Hopkins, P.~F.\ 2016, \mnras, 458, 673 

\bibitem[Haardt \& Madau(2012)]{haardt_2012} Haardt, F., \& Madau, P.\ 2012,
\apj, 746, 125 

\bibitem[Madau \& Haardt(2015)]{haardt_2015} Madau, P., \& Haardt, F.\ 2015,
\apjl, 813, L8 

\bibitem[Hathi et al.(2008)]{hathi_2008} Hathi, N.~P., Jansen, R.~A., Windhorst,
R.~A., et al.\ 2008, \aj, 135, 156 

\bibitem[{{Helgason} {et~al.}(2016){Helgason}, {Ricotti}, {Kashlinsky}, \&
{Bromm}}]{helgason_2016} {Helgason}, K., {Ricotti}, M., {Kashlinsky}, A., \&
{Bromm}, V. 2016, \mnras, 455, 282

\bibitem[Henze et al.(2015)]{henze_2015} Henze, M., Ness, J.-U., Darnley,
M.~J., et al.\ 2015, \aap, 580, A46 

\bibitem[HESS~Collaboration(2013)]{abramowski_2013} HESS~Collaboration,
Abramowski, A., Acero, F., et al.\ 2013, \aap, 550, A4

\bibitem[H.~E.~S.~S.~Collaboration et al.(2017)]{abdalla_2017} 
H.~E.~S.~S.~Collaboration, Abdalla, H., Abramowski, A., et al.\ 2017, \aap, 606,
A59 

\bibitem[Hinshaw et al.(2009)]{hinshaw_2009} Hinshaw, G., Weiland, J.~L., Hill,
R.~S., et al.\ 2009, \apjs, 180, 225 

\bibitem[{{Hirschi}(2007)}]{hirschi_2007} {Hirschi}, R. 2007, \aap, 461, 571

\bibitem[Hoffman et al.(2015)]{hoffman_2015} Hoffman, Y., Courtois, H.~M., \&
Tully, R.~B.\ 2015, \mnras, 449, 4494 

\bibitem[Hoffman et al.(2017)]{hoffman_2017} Hoffman, Y., Pomar{\`e}de, D.,
Tully, R.~B., \& Courtois, H.~M.\ 2017, Nature Astronomy, 1, 0036 

\bibitem[Hogg(1999)]{hogg_1999} Hogg, D.~W.\ 1999, astro-ph/9905116 

\bibitem[Hogg et al.(2002)]{hogg_2002} Hogg, D.~W., Baldry, I.~K., Blanton,
M.~R., \& Eisenstein, D.~J.\ 2002, astro-ph/0210394 

\bibitem[Hosokawa et al.(2016)]{hosokawa_2016} Hosokawa, T., Hirano, S.,
Kuiper, R., et al.\ 2016, \apj, 824, 119 

\bibitem[Hoyle \& Lyttleton(1942)]{hoyle_1942} Hoyle, F., Lyttleton, R.A. \
1942, \mnras, 102, 177

\bibitem[{Hunter(2007)}]{hunter_2007} Hunter, J.~D. 2007, Computing In Science
\& Engineering, 9, 90 (doi:10.1109/MCSE.2007.55)

\bibitem[Ishiyama et al.(2016)]{ishiyama_2016} Ishiyama, T., Sudo, K., Yokoi,
S., et al.\ 2016, \apj, 826, 9 

\bibitem[Jansen et al.(2017)]{jansen_2017} Jansen, R.~A., \& Webb Medium Deep
Fields IDS GTO team 2017, American Astronomical Society Meeting Abstracts
\#229, 438.04 


\bibitem[Jauzac et al.(2014)]{jauzac_2014} Jauzac, M., Cl{\'e}ment, B.,
Limousin, M., et al.\ 2014, \mnras, 443, 1549 

\bibitem[Jauzac et al.(2015)]{jauzac_2015} Jauzac, M., Richard, J., Jullo, E.,
et al.\ 2015, \mnras, 452, 1437 

\bibitem[Jiang et al.(2007)]{jiang_2007} Jiang, L., Fan, X., Vestergaard, M.,
et al.\ 2007, \aj, 134, 1150 

\bibitem[Kashlinsky et al.(2012)]{kashlinsky_2012} Kashlinsky, A., Arendt,
R.~G., Ashby, M.~L.~N., et al.\ 2012, \apj, 753, 63 

\bibitem[Kashlinsky et al.(2015)]{kashlinsky_2015} Kashlinsky, A., Mather,
J.~C., Helgason, K., et al.\ 2015, \apj, 804, 99 

\bibitem[Kashlinsky(2016)]{kashlinsky_2016} Kashlinsky, A.\ 2016, \apjl, 823,
L25

\bibitem[Kaurov et al.(2019)]{kaurov_2019} Kaurov, A.~A., Dai, L., Venumadhav,
T., Miralda-Escud{\'e}, J., \& Frye, B.\ 2019, astro-ph/1902.10090

\bibitem[Kayser et al.(1986)]{kayser_1986} Kayser, R., Refsdal, S., \& Stabell,
R.\ 1986, \aap, 166, 36 

\bibitem[Kawamata et al.(2016)]{kawamata_2016} Kawamata, R., Oguri, M.,
Ishigaki, M., Shimasaku, K., \& Ouchi, M.\ 2016, \apj, 819, 114 

\bibitem[Kelly et al.(2017)]{kelly_2017} Kelly, P.~L., Diego, J.~M.,
Nonino, M., et al.\ 2017, The Astronomer's Telegram, 10005,
(http://adsabs.harvard.edu/abs/2017ATel10005....1K)

\bibitem[Kelly et al.(2018)]{kelly_2018} Kelly, P.~L., Diego, J.~M., Rodney, 
S., et al.\ 2018, Nature Astr., 2, 334 

\bibitem[Kelsall et al.(1998)]{kelsall_1998} Kelsall, T., Weiland, J.~L., Franz,
B.~A., et al.\ 1998, \apj, 508, 44 

\bibitem[Kennicutt(1998)]{kennicutt_1998} Kennicutt, R.~C., Jr.\ 1998, \apj,
498, 541 

\bibitem[Kim et al.(2017)]{kim_2017} Kim, D., Jansen, R.~A., \& Windhorst,
R.~A.\ 2017, \apj, 804, 28 

\bibitem[{{Kiminki} \& {Kobulnicky}(2012)}]{kiminki_2012} {Kiminki}, D.~C., \&
{Kobulnicky}, H.~A. 2012, \apj, 751, 4

\bibitem[Koekemoer et al.(2011)]{koekemoer_2011} Koekemoer, A.~M., Faber,
S.~M., Ferguson, H.~C., et al.\ 2011, \apjs, 197, 36 

\bibitem[Koekemoer et al.(2013)]{koekemoer_2013} Koekemoer, A.~M., Ellis, R.~S.,
McLure, R.~J., et al.\ 2013, \apjs, 209, 3 

\bibitem[Kohri et al.(2014)]{kohri_2014} Kohri, K., Nakama, T., \& Suyama,
T.\ 2014, \prd, 90, 083514 

\bibitem[Koz{\l}owski et al.(2010)]{kozlowski_2010} Koz{\l}owski, S.,
Kochanek, C.~S., Udalski, A., et al.\ 2010, \apj, 708, 927 

\bibitem[{{Kozyreva} \& {Blinnikov}(2015)}]{kozyreva_2015} {Kozyreva}, A., \&
{Blinnikov}, S. 2015, \mnras, 454, 4357

\bibitem[{{Kozyreva} {et~al.}(2017){Kozyreva}, {Gilmer}, {Hirschi},
{Fr{\"o}hlich}, {Blinnikov}, {Wollaeger}, {Noebauer}, {van Rossum}, {Heger},
{Even}, {Waldman}, {Tolstov}, {Chatzopoulos}, \& {Sorokina}}]{kozyreva_2017}
{Kozyreva}, A., {Gilmer}, M., {Hirschi}, R., {et~al.} 2017, \mnras, 464, 2854

\bibitem[Kurk et al.(2007)]{kurk_2007} Kurk, J.~D., Walter, F., Fan,
X., et al.\ 2007, \apj, 669, 32 

\bibitem[Kurucz(2005)]{kurucz_2005} Kurucz, R.~L.\ 2005, Mem. S.A.It. Suppl.,
8, 189 \url{http://kurucz.harvard.edu/sun.html}

\bibitem[Lagattuta et al.(2017)]{lagattuta_2017} Lagattuta, D.~J., Richard,
J., Cl{\'e}ment, B., et al.\ 2017, \mnras, 469, 3946 

\bibitem[Lam et al.(2014)]{lam_2014} Lam, D., Broadhurst, T., Diego, J.~M., et
al.\ 2014, \apj, 797, 98 

\bibitem[{{Lewis} {et~al.}(2000){Lewis}, {Ibata}, \& {Wyithe}}]{lewis_2000}
{Lewis}, G.~F., {Ibata}, R.~A., \& {Wyithe}, J.~S.~B. 2000, \apjl, 542, L9

\bibitem[Livermore et al.(2017)]{livermore_2017} Livermore, R.~C., Finkelstein,
S.~L., \& Lotz, J.~M.\ 2017, \apj, 835, 113 

\bibitem[Lorentz et al.(2015)]{lorentz_2015} Lorentz, M., Brun, P., \&
Sanchez, D.\ 2015, in Proc. of the 34th International Cosmic Ray Conference 
(ICRC2015), Eds. A. M. van den Berg et al.\ (The Hague, The Netherlands:
Proceedings of Science), 34, 777 
(https://pos.sissa.it/cgi-bin/reader/conf.cgi?confid=236)

\bibitem[Lotz et al.(2017)]{lotz_2017} Lotz, J.~M., Koekemoer, A., Coe, D., et
al.\ 2017, \apj, 837, 97 

\bibitem[Machida et al.(2009)]{machida_2009} Machida, M.~N., Omukai, K.,
Matsumoto, T., \& Inutsuka, S.-I.\ 2009, \mnras, 399, 1255 

\bibitem[{{Macpherson} {et~al.}(2013){Macpherson}, {Coward}, \&
{Zadnik}}]{macpherson_2013} {Macpherson}, D., {Coward}, D.~M., \& {Zadnik},
M.~G. 2013, \apj, 779, 73

\bibitem[{{Madau} \& {Silk}(2005)}]{madau_2005} {Madau}, P., \& {Silk}, J. 2005,
\mnras, 359, L37

\bibitem[Madau \& Dickinson(2014)]{madau_2014} Madau, P., \& Dickinson, M.\
2014, \araa, 52, 415 

\bibitem[Madau \& Fragos(2017)]{madau_2017} Madau, P., \& Fragos, T.\ 2017,
\apj, 840, 39 

\bibitem[Mahler et al.(2018)]{mahler_2018} Mahler, G., Richard, J.,
Cl{\'e}ment, B., et al.\ 2018, \mnras, 473, 663 

\bibitem[Maiolino et al.(2008)]{maiolino_2008} Maiolino, R., Nagao, T., Grazian,
A., et al.\ 2008, \aap, 488, 463 

\bibitem[Mamajek et al.(2015)]{mamajek_2015} Mamajek, E.~E., Prsa, A., Torres,
G., et al.\ 2015, astro-ph/1510.07674 

\bibitem[Mas-Ribas et al.(2016)]{mas-ribas_2016} Mas-Ribas, L., Dijkstra, M., \&
Forero-Romero, J.~E.\ 2016, \apj, 833, 65 

\bibitem[Matsuoka et al.(2011)]{matsuoka_2011} Matsuoka, Y., Ienaka, N., Kawara,
K., \& Oyabu, S.\ 2011, \apj, 736, 119 

\bibitem[Mattila et al.(2017)]{mattila_2017} Mattila, K., V{\"a}is{\"a}nen, P.,
Lehtinen, K., von Appen-Schnur, G., \& Leinert, C.\ 2017, \mnras, 470, 2152 

\bibitem[Mayer et al.(2017)]{mayer_2017} Mayer, P., Harmanec, P., Chini, R., et
al.\ 2017, \aap, 600, A33 

\bibitem[Meneghetti et al.(2017)]{meneghetti_2017} Meneghetti, M., Natarajan,
P., Coe, D., et al.\ 2017, \mnras, 472, 3177 

\bibitem[Milosavljevi{\'c} et al.(2009)]{milosavljevic_2009} Milosavljevi{\'c},
M., Bromm, V., Couch, S.~M., \& Oh, S.~P.\ 2009, \apj, 698, 766 

\bibitem[Miralda-Escude(1991)]{miralda_escude_1991} Miralda-Escude, J.\ 1991,
\apj, 379, 94 

\bibitem[{{Mitchell-Wynne} {et~al.}(2016){Mitchell-Wynne}, {Cooray}, {Xue},
{Luo}, {Brandt}, \& {Koekemoer}}]{mitchell-wynne_2016} {Mitchell-Wynne}, K.,
{Cooray}, A., {Xue}, Y., {et~al.} 2016, \apj, 832, 104

\bibitem[Molnar et al.(2013)]{molnar_2013} Molnar, S.~M., Broadhurst, T.,
Umetsu, K., et al.\ 2013, \apj, 774, 70 

\bibitem[Morishita et al.(2017)]{morishita_2017} Morishita, T., Abramson,
L.~E., Treu, T., et al.\ 2017, \apj, 846, 139 

\bibitem[Morgan et al.(2015)]{morgan_2015} Morgan, R.~J., Windhorst, R.~A.,
Scannapieco, E., \& Thacker, R.~J.\ 2015, \pasp, 127, 803 

\bibitem[Mortlock et al.(2011)]{mortlock_2011} Mortlock, D.~J., Warren, S.~J.,
Venemans, B.~P., et al.\ 2011, \nat, 474, 616 

\bibitem[Natarajan et al.(2017)]{natarajan_2017} Natarajan, P., Chadayammuri,
U., Jauzac, M., et al.\ 2017, \mnras, 468, 1962 

\bibitem[Negrello et al.(2017)]{negrello_2017} Negrello, M., Amber, S.,
Amvrosiadis, A., et al.\ 2017, \mnras, 465, 3558 

\bibitem[Oguri et al.(2018)]{oguri_2017} Oguri, M., Diego, J.~M., Kaiser, N.,
Kelly, P.~L., \& Broadhurst, T.\ 2018, \prd, 97, 023518 

\bibitem[{{Ohkubo} {et~al.}(2009){Ohkubo}, {Nomoto}, {Umeda}, {Yoshida}, \&
{Tsuruta}}]{ohkubo_2009} {Ohkubo}, T., {Nomoto}, K., {Umeda}, H., {Yoshida}, N.,
\& {Tsuruta}, S. 2009, \apj, 706, 1184

\bibitem[Oke \& Gunn(1983)]{okegunn_1983} Oke, J.~B., \& Gunn, J.~E.\ 1983,
\apj, 266, 713 

\bibitem[Owers et al.(2011)]{owers_2011} Owers, M.~S., Randall, S.~W., Nulsen,
P.~E.~J., et al.\ 2011, \apj, 728, 27 

\bibitem[Pagel \& Portinari(1998)]{pagel_1998} Pagel, B.E.J., \& Portinari, L.\
1998, \mnras, 298, 747

\bibitem[Park \& Ricotti(2012)]{park_2012} Park, K., \& Ricotti, M.\ 2012, \apj,
747, 9 

\bibitem[{{Paxton} {et~al.}(2011){Paxton}, {Bildsten}, {Dotter}, {Herwig},
{Lesaffre}, \& {Timmes}}]{paxton_2011} {Paxton}, B., {Bildsten}, L., {Dotter},
A., {et~al.} 2011, \apjs, 192, 3

\bibitem[{{Paxton} {et~al.}(2013){Paxton}, {Cantiello}, {Arras}, {Bildsten},
{Brown}, {Dotter}, {Mankovich}, {Montgomery}, {Stello}, {Timmes}, \&
{Townsend}}]{paxton_2013} {Paxton}, B., {Cantiello}, M., {Arras}, P., {et~al.}
2013, \apjs, 208, 4

\bibitem[{{Paxton} {et~al.}(2015){Paxton}, {Marchant}, {Schwab}, {Bauer},
{Bildsten}, {Cantiello}, {Dessart}, {Farmer}, {Hu}, {Langer}, {Townsend},
{Townsley}, \& {Timmes}}]{paxton_2015} {Paxton}, B., {Marchant}, P., {Schwab},
J., {et~al.} 2015, \apjs, 220, 15

\bibitem[Petermann \& Timmes(2018)]{petermann_2018} Petermann, I., \& Timmes, F.
X. 2018, private communication

\bibitem[Planck Collaboration et al.(2014)]{planck_XXVII_2014} Planck
Collaboration, Aghanim, N., Armitage-Caplan, C., et al.\ 2014, \aap, 571, A27 

\bibitem[Planck Collaboration et al.(2016{\natexlab{a}})]{planck_XIII_2016_a}
Planck Collaboration, Ade, P.~A.~R., Aghanim, N., et al.\ 2016{\natexlab{a}},
\aap, 594, A13 

\bibitem[Planck Collaboration et al.(2016{\natexlab{b}})]{planck_XLVI_2016_b}
Planck Collaboration, Aghanim, N., Ashdown, M., et al.\ 2016{\natexlab{b}},
\aap, 596, A107 

\bibitem[Planck Collaboration et al.(2016{\natexlab{c}})]{planck_II_2016_c}
Planck Collaboration, Ade, P.~A.~R., Aghanim, N., et al.\ 2016{\natexlab{c}},
\aap, 594, A2 

\bibitem[Planck Collaboration et al.(2016{\natexlab{d}})]{planck_XLVII_2016_d}
Planck Collaboration, Adam, R., Aghanim, N., et al.\ 2016{\natexlab{d}}, \aap,
596, A108 

\bibitem[Portinari et al.(2010)]{portinari_2010} Portinari, L., Casagrande, L.,
\& Flynn, C.\ 2010, \mnras, 406, 1570 

\bibitem[Postman et al.(2012)]{postman_2012} Postman, M., Coe, D.,
Ben{\'{\i}}tez, N., et al.\ 2012, \apjs, 199, 25 

\bibitem[Pr{\v s}a et al.(2016)]{prsa_2016} Pr{\v s}a, A., Harmanec, P., Torres,
G., et al.\ 2016, \aj, 152, 41 

\bibitem[Remillard \& McClintock(2006)]{remillard_2006} Remillard, R.~A., \&
McClintock, J.~E.\ 2006, \araa, 44, 49 

\bibitem[Renzo et al.(2017)]{renzo_2017} Renzo, M., Ott, C.~D., Shore, S.~N.,
\& de Mink, S.~E.\ 2017, \aap, 603, A118 

\bibitem[Rieke et al.(2005)]{rieke_2005} Rieke, M.~J., Kelly, D., \& Horner,
S.\ 2005, \procspie, 5904, 1 

\bibitem[Robotham et al.(2011)]{robotham_2011} Robotham, A.~S.~G., Norberg, P.,
Driver, S.~P., et al.\ 2011, \mnras, 416, 2640 

\bibitem[Rodney et al.(2018)]{rodney_2018} Rodney, S.~A., Balestra, I.,
Bradac, M., et al.\ 2018, Nature Astr., 2, 324 

\bibitem[Romero et al.(2015)]{romero_2015} Romero, A.~D., Campos, F., \&
Kepler, S.~O.\ 2015, \mnras, 450, 3708 

\bibitem[{{Rydberg} {et~al.}(2013){Rydberg}, {Zackrisson}, {Lundqvist}, \&
{Scott}}]{rydberg_2013} {Rydberg}, C.-E., {Zackrisson}, E., {Lundqvist}, P., \&
{Scott}, P. 2013, \mnras, 429, 3658

\bibitem[Rydberg et al.(2015)]{rydberg_2015} Rydberg, C.-E., Zackrisson, E.,
Zitrin, A., et al.\ 2015, \apj, 804, 13 

\bibitem[Salpeter(1955)]{salpeter_1955} Salpeter, E.~E.\ 1955, \apj, 121, 161 

\bibitem[{{Sana} {et~al.}(2012){Sana}, {de Mink}, {de Koter}, {Langer}, {Evans},
{Gieles}, {Gosset}, {Izzard}, {Le Bouquin}, \& {Schneider}}]{sana_2012} {Sana},
H., {de Mink}, S.~E., {de Koter}, A., {et~al.} 2012, Science, 337, 444

\bibitem[{{Sarmento} {et~al.}(2017){Sarmento}, {Scannapieco}, \&
{Pan}}]{sarmento_2017} {Sarmento}, R., {Scannapieco}, E., \& {Pan}, L. 2017,
\apj, 834, 23

\bibitem[Sarmento et al.(2018)]{sarmento_2018} Sarmento, R., Scannapieco, E.,
\& Cohen, S.\ 2018, \apj, 854, 75 

\bibitem[Sarmento et al.(2019)]{sarmento_2019} Sarmento, R., Scannapieco, E.,
\& C{\^o}t{\'e}, B.\ 2019, \apj, 871, 206 

\bibitem[Scalo(1986)]{scalo_1986} Scalo, J.~M.\ 1986, \fcp, 11, 1 
(http://adsabs.harvard.edu/abs/1986FCPh...11....1S)

\bibitem[Schaerer(2002)]{schaerer_2002} Schaerer, D.\ 2002, \aap, 382, 28 

\bibitem[Shafter et al.(2015)]{shafter_2015} Shafter, A.~W., Henze, M.,
Rector, T.~A., et al.\ 2015, \apjs, 216, 34 

\bibitem[Shafter(2017)]{shafter_2017} Shafter, A.~W.\ 2017, \apj, 834, 196 

\bibitem[Shakura \& Sunyaev(1973)]{shakura_sunyaev_1973} Shakura, N.~I., \&
Sunyaev, R.~A.\ 1973, \aap, 24, 337 

\bibitem[Shakura \& Sunyaev(1976)]{shakura_sunyaev_1976} Shakura, N.~I., \&
Sunyaev, R.~A.\ 1976, \mnras, 175, 613 

\bibitem[Shara et al.(1986)]{shara_1986} Shara, M.~M., Livio, M., Moffat,
A.~F.~J., \& Orio, M.\ 1986, \apj, 311, 163 

\bibitem[{{Smith} {et~al.}(2007){Smith}, {Li}, {Foley}, {Wheeler}, {Pooley},
{Chornock}, {Filippenko}, {Silverman}, {Quimby}, {Bloom}, \&
{Hansen}}]{smith_2007} {Smith}, N., {Li}, W., {Foley}, R.~J., {et~al.} 2007,
\apj, 666, 1116

\bibitem[Smith et al.(2018)]{smith_2018} Smith, B.~M., Windhorst, R.~A.,
Jansen, R.~A., et al.\ 2018, \apj, 853, 191 

\bibitem[Sobral et al.(2015)]{sobral_2015} Sobral, D., Matthee, J., Darvish, B.,
et al.\ 2015, \apj, 808, 139

\bibitem[Springel \& Farrar(2007)]{springel_2007} Springel, V., \& Farrar,
G.~R.\ 2007, \mnras, 380, 911 

\bibitem[Stacy et al.(2016)]{stacy_2016} Stacy, A., Bromm, V., \& Lee, A.~T.\
2016, \mnras, 462, 1307 

\bibitem[Stanway et al.(2016)]{stanway_2016} Stanway, E.~R., Eldridge, J.~J., \&
Becker, G.~D.\ 2016, \mnras, 456, 485 

\bibitem[{{Sugimoto} \& {Nomoto}(1980)}]{sugimoto_1980} {Sugimoto}, D., \&
{Nomoto}, K. 1980, \ssr, 25, 155

\bibitem[Sukhbold \& Woosley(2014)]{sukhbold_2014} Sukhbold, T., \& Woosley,
S.~E.\ 2014, \apj, 783, 10 

\bibitem[Sukhbold \& Woosley(2016)]{sukhbold_2016} Sukhbold, T., \& Woosley,
S.~E.\ 2016, \apjl, 820, L38 

\bibitem[Susa et al.(2014)]{susa_2014} Susa, H., Hasegawa, K., \& Tominaga, N.\
2014, \apj, 792, 32 

\bibitem[Tanaka et al.(2012)]{tanaka_2012} Tanaka, T., Perna, R., \& Haiman, Z.\
2012, \mnras, 425, 2974 

\bibitem[Tanaka \& Shibazaki(1996)]{tanaka_1996} Tanaka, Y., \& Shibazaki, N.\
1996, \araa, 34, 607 

\bibitem[Thompson \& Nagamine(2012)]{thompson_2012} Thompson, R., \& Nagamine,
K.\ 2012, \mnras, 419, 3560 

\bibitem[{{Trenti} \& {Stiavelli}(2007)}]{trenti_2007} {Trenti}, M., \&
{Stiavelli}, M. 2007, \apj, 667, 38

\bibitem[{{Trenti} \& {Stiavelli}(2009)}]{trenti_2009} ---. 2009, \apj, 694, 879

\bibitem[Trujillo \& Fliri(2016)]{trujillo_2016} Trujillo, I., \& Fliri, J.\
2016, \apj, 823, 123 

\bibitem[Tucker et al.(1998)]{tucker_1998} Tucker, W., Blanco, P., Rappaport,
S., et al.\ 1998, \apjl, 496, L5 

\bibitem[Turk et al.(2009)]{turk_2009} Turk, M.~J., Abel, T., \& O'Shea, B.\
2009, Science, 325, 601 

\bibitem[{van~der Walt {et~al.}(2011)van~der Walt, Colbert, \&
Varoquaux}]{der_walt_2011} van~der Walt, S., Colbert, S.~C., \& Varoquaux, G.
2011, Computing in Science Engineering, 13, 22 (doi:10.1109/MCSE.2011.37)

\bibitem[Watkins \& Feldman(2015{\natexlab{a}})]{watkins_2015_a} Watkins, R., \&
Feldman, H.~A.\ 2015{\natexlab{a}}, \mnras, 447, 132 

\bibitem[Watkins \& Feldman(2015{\natexlab{b}})]{watkins_2015_b} Watkins, R., \&
Feldman, H.~A.\ 2015{\natexlab{b}}, \mnras, 450, 1868 

\bibitem[Watts(2012)]{watts_2012} Watts, A.~L.\ 2012, \araa, 50, 609 

\bibitem[Watson et al.(2014)]{watson_2014} Watson, W.~A., Iliev, I.~T., Diego,
J.~M., et al.\ 2014, \mnras, 437, 3776 

\bibitem[{{Wheeler}(1977)}]{wheeler_1977} {Wheeler}, J.~C. 1977, \apss, 50, 125

\bibitem[Willott et al.(2003)]{willott_2003} Willott, C.~J., McLure,
R.~J., \& Jarvis, M.~J.\ 2003, \apjl, 587, L15 

\bibitem[Willott et al.(2010)]{willott_2010} Willott, C.~J., Albert, L.,
Arzoumanian, D., et al.\ 2010, \aj, 140, 546 

\bibitem[{{Windhorst} {et~al.}(2008){Windhorst}, {Hathi}, {Cohen}, {Jansen},
{Kawata}, {Driver}, \& {Gibson}}]{windhorst_2008} {Windhorst}, R.~A., {Hathi},
N.~P., {Cohen}, S.~H., {et~al.} 2008, Advances in Space Research, 41, 1965 

\bibitem[{{Windhorst} {et~al.}(2011){Windhorst}, {Cohen}, {Hathi}, {McCarthy},
{Ryan}, {Yan}, {Baldry}, {Driver}, {Frogel}, {Hill}, {Kelvin}, {Koekemoer},
{Mechtley}, {O'Connell}, {Robotham}, {Rutkowski}, {Seibert}, {Straughn},
{Tuffs}, {Balick}, {Bond}, {Bushouse}, {Calzetti}, {Crockett}, {Disney},
{Dopita}, {Hall}, {Holtzman}, {Kaviraj}, {Kimble}, {MacKenty}, {Mutchler},
{Paresce}, {Saha}, {Silk}, {Trauger}, {Walker}, {Whitmore}, \&
{Young}}]{windhorst_2011} {Windhorst}, R.~A., {Cohen}, S.~H., {Hathi}, N.~P.,
{et~al.} 2011, \apjs, 193, 27 (W11)

\bibitem[Windhorst et al.(2018)]{windhorst_2018} Windhorst, R.~A., Timmes,
F.~X., Wyithe, J.~S.~B., et al.\ 2018, \apjs, 234, 41 (W18; astro-ph/1901.00565)

\bibitem[Wolf et al.(2013)]{wolf_2013} Wolf, W.~M., Bildsten, L., Brooks, J.,
\& Paxton, B.\ 2013, \apj, 777, 136 

\bibitem[{Woosley {et~al.}(2002)Woosley, Heger, \& Weaver}]{woosley_2002}
Woosley, S.~E., Heger, A., \& Weaver, T.~A. 2002, Rev. Mod. Phys., 74, 1015

\bibitem[Woosley(2017)]{woosley_2017} Woosley, S.~E.\ 2017, \apj, 836, 244 

\bibitem[{{Yoon} {et~al.}(2008){Yoon}, {Cantiello}, \& {Langer}}]{yoon_2008}
{Yoon}, S.-C., {Cantiello}, M., \& {Langer}, N. 2008, in American Institute of
Physics Conference Series, Vol. 990, First Stars III, ed. B.~W. {O'Shea} \&
A.~{Heger}, 225--229

\bibitem[Yue et al.(2013)]{yue_2013} Yue, B., Ferrara, A., Salvaterra, R., Xu,
Y., \& Chen, X.\ 2013, \mnras, 433, 1556 

\bibitem[{{Yusof} {et~al.}(2013){Yusof}, {Hirschi}, {Meynet}, {Crowther},
{Ekstr{\"o}m}, {Frischknecht}, {Georgy}, {Abu Kassim}, \&
{Schnurr}}]{yusof_2013} {Yusof}, N., {Hirschi}, R., {Meynet}, G., {et~al.}
2013, \mnras, 433, 1114

\bibitem[{{Zackrisson} {et~al.}(2015){Zackrisson}, {Gonz{\'a}lez}, {Eriksson},
{Asadi}, {Safranek-Shrader}, {Trenti}, \& {Inoue}}]{zackrisson_2015}
{Zackrisson}, E., {Gonz{\'a}lez}, J., {Eriksson}, S., {et~al.} 2015, \mnras,
449, 3057

\bibitem[Zemcov et al.(2017)]{zemcov_2017} Zemcov, M., Immel, P., Nguyen, C.,
et al.\ 2017, Nature Communications, 8, 15003 

\bibitem[Zhang et al.(2010)]{zhang_2010} Zhang, F., Han, Z., Li, L., Guo, J., \&
Zhang, Y.\ 2010, \apss, 329, 249 

\bibitem[Zitrin et al.(2013)]{zitrin_2013} Zitrin, A., Menanteau, F., Hughes,
J.~P., et al.\ 2013, \apjl, 770, L15 

\end{thebibliography}

\ve 


\end{document}